\documentclass[aps,prd,twocolumn,showpacs,nofootinbib]{revtex4}
\usepackage{amsmath}
\usepackage{graphicx}
\usepackage{dcolumn}
\usepackage{bm}
\usepackage{amssymb}
\usepackage{latexsym}
\usepackage{bm}

\usepackage{float}

\newcommand{\no}{\nonumber}

\newcommand{\pa}{\partial}
\newcommand{\del}{\delta}
\newcommand{\eps}{\epsilon}

{\bf }

\newcommand{\half}{\frac{1}{2}}

\begin{document}


\title{Simulate Anyons by cold  atoms with  induced electric dipole moment 
}

\author{Jian Jing ${}^{a}$}
\email{jingjian@mail. buct. edu. cn}

\author{Yao-Yao Ma  ${}^{a}$}
\author{Qiu-Yue Zhang ${}^{a}$}

\author{Qing Wang ${}^{b}$}



\author{Shi-Hai Dong ${}^c$}
\email{dongsh2@yahoo. com}

\affiliation{${}^a$ Department of Physics and Electronic, School of
Science, Beijing University of Chemical Technology, Beijing 100029,
P. R. China, }

\affiliation{${}^b$ College of Physics and Technology, Xinjiang University, Urumqi 830046, P. R. China, }


\affiliation{{ $^{c}$ Laboratorio de Informaci\'on Cu\'antica, CIDETEC, Instituto Polit\'{e}cnico Nacional, UPALM, CDMX 07700, M{e}xico. }}


\begin{abstract}

We show that it is possible to simulate an anyon by  a trapped atom  which possesses  an induced electric dipole moment in the background of electromagnetic fields with a specific configuration.  The electromagnetic fields we applied contain a magnetic  and two electric fields. We find that  when the atom  is cooled down to the limit of the negligibly small kinetic energy, the atom behaves like an anyon because its angular momentum  takes fractional values. The fractional part of the angular momentum is determined by both  the magnetic  and one of the electric fields. Roles  two electromagnetic fields  played are analyzed. 

\end{abstract}

{}

\maketitle

\section{Introduction}

Simulation of physical  phenomena which occurs originally in charged particles by neutral ones is an interesting subject. An example is the simulation of Aharononv-Bohm (AB) effect  by neutral particles. AB effect  predicts that  a charged particle will accumulate a geometrical phase when it moves around  a long-thin magnetic-flux carried solenoid \cite{AB}. 

The simulation of AB effect by a neutral particle which possesses a permanent magnetic dipole moment was proposed by Aharonov and Casher.  In Ref. \cite{AC}, they predicted that a neutral particle with a permanent magnetic dipole moment would  acquire a geometrical phase if it moved around a uniformly electric charged long filament with the direction of the magnetic dipole moment paralleling to the filament. It is the Aharonov-Casher (AC) effect. 

The simulation of AB effect by a neutral particle with permanent electric dipole moment was proposed in refs. \cite{HM, Wilkens}.  It was predicted that a neutral particle with a permanent electric dipole moment would receive a geometrical phase if it circled around a uniformly magnetic charged long filament.  It is named He-Mckellar-Wilkens   (HMW) effect.  The  observation of HMW effect in experiments is  difficult since the magnetic field in HMW effect is produced by magnetic monopoles \cite{dwf, wilkens2}.   

In order to avoid this difficulty, Ref. \cite{han} proposed an alternative method to observe the HMW effect. Instead of using a neutral particle which possesses a permanent electric dipole moment, the authors of this paper proposed to use a neutral particle with an induced electric dipole moment interacting with an electric and a magnetic fields.  Compared with HMW effect, the magnetic field in the proposal \cite{han} is easily prepared in experiments. 

Another  example  is Landau levels. Landau levels are eigenvalues of a charged planar particle interacting with a uniform perpendicular magnetic field. In Ref. \cite{es}, the authors  showed that  Landau levels could be simulated by an atom which possesses a permanent magnetic dipole moment in the background of an electric field.  Since then, there are many research work concerning   the analogy between  Landau levels and  spectra of  neutral particles which possess  permanent electric or magnetic dipoles interacting with electromagnetic fields \cite{RFN, FNR, Banerjee, Bakke1, Bakke2, Basu, bakke1, bakke2, bakke3, bakke4, bakke5, bakke6, bakke7}. 


We shall show that anyons \cite{anyons1, anyons2}, which was mostly realized by charged particles before, can also be simulated by a neutral particle with an induced electric dipole moment. As is known, eigenvalues of the canonical angular momentum must be quantized  in the three-dimensional space  \cite{Dirac1, Sakurai}. However, in the two-dimensional space, eigenvalues of the canonical angular momentum  can take fractional values \cite{wilczek1, liang}.  The reason is that  the rotation group in three-dimensional space is a non-Abelian one while it is Abelian in the two-dimensional space. 
Particles which have the fractional angular momentum (FAM) are named  anyons \cite{anyons1, anyons2}. Anyons play important roles in  understanding  quantum Hall effects \cite{hall} and high $T_c$ superconductivity \cite{wilczek}. There are several ways to realize anyons. One of the ways is to  couple a charged particle  to Chern-Simons gauge field in ($2+1$)-dimensional space-time \cite{cs1, cs2, cs3, forte, note}.  Recently, anyons receive renewed interests \cite{yhzhang, yhzhang2, dlnr}.

Ref.  \cite{zhang}  proposed an alternative  approach to realize anyons. The author of this reference  coupled an ion to two  magnetic fields. One  is a uniform magnetic field and the other is generated by a long-thin magnetic solenoid. Provided the kinetic energy of this ion is cooled down to its lowest level by using the cold atomic technologies, the author found that eigenvalues of the canonical angular momentum of this charged particle can take fractional values. The fractional part  is determined by the magnetic flux inside the magnetic solenoid.  


In this paper, we propose to simulate  anyons by coupling neutral particles, for example, atoms, which possesses an induced electric dipole moment to electromagnetic fields. The electromagnetic fields we applied  contain a magnetic field and two electric fields. The organization of this paper is as follows: in next section, we introduce our model. Then, we quantize the model canonically and pay attention to its rotation property. Although the canonical angular momentum of this model only can take integer values, we show that the canonical angular momentum of the reduced model, which is obtained by cooling down the kinetic energy of the atom to the negligibly small, takes fractional values. The fractional part of the angular momentum depends on the intensity of the magnetic and only one of
the electric fields explicitly. In section III, we analyze the roles two electric fields played in the simulation of anyons. We prove that both of the electric fields are  necessary to simulate anyons. Summations and conclusions will be given in the last section.

\section{Fractional angular momentum}

The model we considered is  an  atom which possesses an induced electric dipole moment  interacting with electromagnetic fields. Electromagnetic fields we applied consist of a pair of electric fields $\mathbf E^{(1)},  \ \mathbf E^{(2)}$ and a uniform magnetic field $\mathbf B$.  
The electric field $\mathbf E^{(1)}$ is produced by a long filament with uniform electric charges per length, $\mathbf E^{(2)}$ is produced by the uniformly distributed electric charges. The magnetic field is along the $z$-direction and electric fields are in the radial direction of the plane which is perpendicular to the magnetic field. 
Explicitly, the electromagnetic fields we considered are
\begin{equation}
\mathbf E^{(1)} = \frac{k}{r}\mathbf {e_r}, \ \ \ \mathbf E^{(2)} = \frac{\rho}{2} r \mathbf{e_r} \label{ee},
\end{equation}
and
\begin{equation}
\mathbf B = B \mathbf {e_z} \label{mf}
\end{equation}
where $k$ and $\rho$ are  parameters  which are characters of these two electric fields, and $\mathbf e_r$ is the unit vector along the radial direction on the plane. Besides the electromagnetic fields (\ref{ee}) and (\ref{mf}), the atom is trapped by a harmonic potential simultaneously. The harmonic potential and 
electromagnetic fields are designed to make the motion of the atom be rotationally symmetric.

In Ref. \cite{han}, the authors showed that  a neutral atom with an induced electric dipole moment would receive a topological phase if it  moves around this uniformly electric charged filament in the presence of the magnetic field (\ref{mf}). 
By confining the atom on a rigid circle, the authors of  \cite{ASCFC} investigated the eigenvalue problem of the model  considered here. 

Due to the electric fields, the atom will be polarized, i.e., it will induce an electric dipole moment (we set $c=1$)
\begin{equation}
\mathbf d= \alpha (\mathbf E + \mathbf v \times \mathbf B)
\end{equation}
where $\alpha,  \ \mathbf v$ are the dielectric polarizability and the velocity of the atom respectively, and $\mathbf E $ is the summation of two electric fields,  $\mathbf E = \mathbf E^{(1)} + \mathbf E^{(2)}$. The second term on the right-hand side of  the above equation actually is the relativistic effect, it reflects the fact that a moving particle in a magnetic field will feel an electric field $\sim \mathbf v \times \mathbf B$ \cite{jackson}.

Taking the electromagnetic fields (\ref{ee}) and (\ref{mf}) into account  and trapping the atom by a harmonic potential, we get the Lagrangian which describes the dynamics of the  atom. It is  
\begin{equation}
L = \half m \mathbf v ^2 + \half \mathbf d \cdot (\mathbf E + \mathbf v \times \mathbf B) - \half K \mathbf r ^2, \label{la1}
\end{equation}
where the last term is the harmonic potential provided by a trap. 
Substituting the expression $\mathbf d = \alpha (\mathbf E + \mathbf v \times \mathbf B)$ into the above Lagrangian and 
confining the motion of the atom in the plane perpendicular to the magnetic field, we simplify the Lagrangian (\ref{la1}) to the form (the Latin index $i, \ j$ take values $1, \ 2$ and the summation convention is applied throughout the present paper)
\begin{equation}
L= \half M \dot x_i ^2 - \alpha B \eps_{ij}\dot x_i E_j + \half \alpha E_i ^2 - \half K x_i ^2, 
\label{la3}
\end{equation}
where $ M = m + \alpha B ^2 $ is the effective mass. 

We should quantize the model (\ref{la3}) before  studying its quantum properties.  To this end, we define the canonical momenta with respect to variables $x_i$,
\begin{equation}
p_i = \frac{\pa L}{\pa \dot x_i} = M \dot x_i - \alpha B \eps_{ij} E_j. \label{cm}
\end{equation}
The classical Poisson brackets among canonical variables $x_i, \ p_i$ are
\begin{equation}
\{x_i, \ x_j \}= \{p_i, \ p_j \}=0, \quad  \{x_i, \ p_j \}= \del_{ij}.  \label{cpb}
\end{equation}
Then the canonical  Hamiltonian is achieved by the Legendre transformation, 
\begin{equation}
H = \frac{1}{2 M} (p_i + \alpha B \eps_{ij} E_j)^2 - \half \alpha E_i ^2 + \half K x_i ^2. \label{ha1}
\end{equation}
The canonical quantization is  accomplished  when the replacements
$$
x_i \to x_i, \ p_i \to - i \hbar \frac{\pa}{\pa x_i}, \ \{\quad, \quad \} \to \frac{1}{i \hbar}[\quad, \ \quad]
$$
in the classical Hamiltonian (\ref{ha1}) and the Poisson brackets (\ref{cpb}) are complete.

The  canonical angular momentum is 
\begin{equation}{}
J = \eps_{ij} x_i p_j, \label{am}
\end{equation}
which is  proved to be conserved, i.e., $[J, \ H]=0$. It can also be written as $J = -i \hbar {\pa}/{\pa \varphi}$ where $\varphi$ is the azimuth angle. Obviously, eigenvalues of this canonical angular momentum must be  quantized, 
\begin{equation}
J_n = n \hbar, \ n=0, \pm 1, \pm 2, \cdots. \label{evam}
\end{equation}

Now, we consider the reduced model which is the limit of taking the  kinetic energy in (\ref{la3}) to be negligibly small. 
This may be realized in experiments by cooling down the atom to a slower velocity so that the effective kinetic energy can be neglected \footnote{In an experiment  carried out in the early of 1990s, the velocity of atoms  can be cooled down to  $\sim 1\ ms^{-1}$ \cite{sst}.}.
This kind of  reduction is first considered in \cite{jackiw} during the studies of the Chern-Simons quantum mechanics. 

The reduced model is described by the Lagrangian 
\begin{equation}
L_r = - \alpha B \eps_{ij} \dot x_i E_j + \half \alpha E_i ^2 - \half K x_i ^2 \label{rl}
\end{equation}
from which we get  canonical momenta with respect to  variables  $x_i$. They are
\begin{equation}
p_i = \frac{\pa L_r}{\pa \dot x_i} = - \alpha B \eps_{ij} E_j. \label{rcm}
\end{equation}
The R.H.S of the above equations does not contain  velocities, thus, they are in fact the primary constraints in the terminology of Dirac \cite{dirac}.  We label them as
\begin{equation}
\phi_i ^{(0)} = p_i + \alpha B \eps_{ij} E_j \approx 0,  \label{constraints}
\end{equation}
in which '$\approx$' means equivalent on the constraint surface. The existence of primary constraints shows that there are dependent  degrees of freedom in the reduced model (\ref{rl}). 
The classical Poisson brackets among these two primary constraints can be obtained by a straightforward calculation. They are
\begin{equation}
\{\phi_i ^{(0)}, \ \phi_j ^{(0)} \} =  \alpha \rho B  \eps_{ij}. \label{pbcon}
\end{equation}
Since $\{\phi_i ^{(0)}, \ \phi_j ^{(0)} \} \neq 0$, the primary constraints $\phi^{(0)}_i$ belong to the second class and there are no secondary constraints. Therefore, the constraints $\phi^{(0)}_i$  can be used to eliminate the dependent degrees of freedom in the reduced model (\ref{rl}). 

The canonical angular momentum in this reduced model has the same expression as (\ref{am}), i.e., $J= \eps_{ij}x_i p_j$. Since there are constraints $\phi^{(0)}_i \approx 0$ which lead to the dependence among canonical variables  $x_i, \ p_i$, we rewrite the canonical angular momentum  by substituting the constraints (\ref{constraints}) into $J= \eps_{ij} x_i p_j$,
\begin{equation}
J= \eps_{ij} x_i p_j = \alpha B x_i E_i.
\end{equation}
Considering the explicit form  of electric field (\ref{ee}), we get
\begin{equation}
J= \alpha B x_i (E_i ^{(1)} + E_i ^{(2)}) = \alpha B (k + \frac{\rho}{2} x_i^2). \label{ram}
\end{equation}
It is more convenient to get eigenvalues of the angular momentum (\ref{ram}) by algebraic method. In doing so, we must determine the commutator between $x_i$ before further proceeding. The classical version of the commutator, i.e., the Dirac bracket, can be calculated by the definition \cite{dirac}.
\begin{equation}
\{x_i, \ x_j \}_D = \{x_i, \ x_j \} - \{ x_i, \ \phi_k ^{(0)} \} \{\phi_k ^{(0)}, \ \phi_l ^{(0)}\}^{-1} \{\phi^{(0)}_l, \ x_j \}.
\end{equation}
Upon straightforward algebraic calculation, we arrive at
\begin{equation}
\{x_i, \ x_j \}_D = -\frac{ \eps_{ij} }{\alpha \rho B }. \label{db1}
\end{equation}
Thus, the commutators between $x_i$ are 
\begin{equation}
[x_i, \ x_j] = -\frac{ i\hbar \eps_{ij} }{\alpha \rho B }.  \label{fqc}
\end{equation}

Taking into account the above commutator,  it is clear to see that apart from the term $\alpha B k$, the canonical angular momentum (\ref{ram}) is equivalent to a one-dimensional harmonic oscillator. With the help of the commutators (\ref{fqc}), one can write down the eigenvalues of the canonical angular momentum (\ref{rcm}) immediately. They are
\begin{equation}
J_n = \alpha B k + (n + \half) \hbar. \label{evram}
\end{equation}

Therefore, it shows that eigenvalues of the canonical angular momentum  will take fractional values when its kinetic energy  is cooled down to the negligibly small. %
The fractional part  is determined by  both the intensity of the applied magnetic field and the electric field 
$\mathbf E^{(1)}$.


From the eigenvalues of the canonical angular momentum (\ref{evram}), it seems that the electric field $\mathbf E^{(2)}$ does not have any influences on  the FAM  since the parameter $\rho$ does not appear in (\ref{evram}) explicitly. In fact, the electric field $\mathbf E^{(2)}$ also plays important roles in producing the FAM. In the next section, we  analyze the roles that  two electric fields $\mathbf E^{(1)}$ and $\mathbf E^{(2)}$ played.  

\section{Roles two electric fields played}

As we showed that besides the intensity of the magnetic field, the fractional part of the canonical angular momentum only contains the parameter $k$. Thus it seems that only the electric field $\mathbf E^{(1)}$ contributes to the FAM. In the following, we show that the electric field $\mathbf E^{(2)}$ also plays important roles in producing the FAM since the FAM  will not appear in the absence of either of  electric fields.  

First of all, we consider the case that  the electric field $\mathbf E^{(1)}$ is turned off. In this case, 
the dynamics is determined by the Lagrangian
\begin{equation}
\bar L= \half M \dot x_i ^2 - \alpha B \eps_{ij}\dot x_i E_j^{(2)} + \half \alpha (E_i^{(2)}) ^2 - \half K x_i ^2. \label{la10}
\end{equation}
Compared with the Lagrangian  (\ref{la3}) in which both of the electric fields are present, we find that the only difference is that the term $E_i= E_i ^{(1)} + E^{(2)}_i$ is replaced by $E_i^{(2)}$. 

The canonical momenta with respective to $x_i$ are given by
\begin{equation}
p_i = \frac{\pa  \bar L}{\pa \dot x_i} = M \dot x_i - \alpha B \eps_{ij} E_j ^{(2)}.
\end{equation}
The  model (\ref{la10}) can be quantized directly.  The canonical angular momentum is defined as usual  $J=\eps_{ij} x_i p_j = -i \hbar \pa /\pa \varphi$ and its eigenvalues are  $J_n =n \hbar, \ 
0, \pm 1, \pm 2, \cdots$. It seems that as far as the rotation property is concerned, there is no difference between the model (\ref{la3}) and the (\ref{la10}).

However, when the atom is cooled down to the negligibly small kinetic energy, their  difference appears. To see it clearly, we set the effective kinetic energy term to zero in  Lagrangian (\ref{la10}) in this limit. Therefore, the Lagrangian (\ref{la10}) reduces to 
\begin{equation}
\bar L _r = - \alpha B \eps_{ij}\dot x_i E_j^{(2)} + \half \alpha (E_i^{(2)}) ^2 - \half K x_i ^2.
\end{equation}
Introducing the canonical momenta with respective to $x_i$, we get two primary constraints as
\begin{equation}
\bar \phi_i ^{(0)} = p_i + \alpha B \eps_{ij} E_j^{(2)} \approx 0,  \label{pc10}
\end{equation}
The Poisson brackets between  constraints (\ref{pc10}) are 
\begin{equation}
\{\bar \phi_i ^{(0)}, \ \bar \phi_j ^{(0)} \} =  \alpha B \rho \eps_{ij} \label{pbcon1}
\end{equation}
which are equivalent to (\ref{pbcon}). Therefore, they are the second class and can be used to eliminate the dependent degrees of freedom. Substituting the constraints (\ref{pc10}) into canonical angular momentum $J=\eps_{ij}x_i p_j$,  we  find that the canonical angular momentum takes the from
\begin{equation}
J=\alpha B x_i  E_i ^{(2)} =   \frac{\alpha  B}{2} x_i^2
\end{equation}
in this limit.
Its eigenvalues  can be obtained once we get the commutators between $x_i$. It can be checked that the Dirac brackets  between $x_i$ are nothing but (\ref{db1}). Thus, eigenvalues of the angular momentum are $J_n = (n+ \half) \hbar, \ n=0, \pm 1, \pm 2, \cdots $.
Therefore, the electric field $\mathbf E^{(2)}$ alone can not produce the FAM.

On the contrary, if we turn off the electric field  $\mathbf E^{(2)}$ and let $\mathbf E^{(1)}$ alone,  the Lagrangian (\ref{la3}) becomes
\begin{equation}
\tilde L= \half M \dot x_i ^2 - \alpha B \eps_{ij}\dot x_i E_j^{(1)} + \half \alpha (E_i^{(1)}) ^2 - \half K x_i ^2. \label{la20}
\end{equation}
We  introduce the canonical momentum 
$
p_i = \frac{\pa \tilde L}{\pa \dot x_i}  = M \dot x_i - \alpha B \eps_{ij} E^{(1)}_j
$
and quantize the model (\ref{la20}) canonically. Then  eigenvalues of the canonical angular momentum $J= \eps_{ij} x_i p_j = -i \hbar \pa /\pa \varphi$ must be quantized as $J_n = n \hbar, \ n=0, \ \pm 1, \ \pm 2, \cdots$. 

The reduced model of the Lagrangian (\ref{la20}) turns out to be 
\begin{equation}
\tilde L_r = - \alpha B \eps_{ij} \dot x_i E_j ^{(1)} + \half \alpha (E_i^{(1)}) ^2 - \half K x_i ^2. \label{rm2}
\end{equation}
The Hamiltonian corresponding to this Lagrangian can be read directly from the above Lagrangian \cite{fj}. It is
\begin{equation}
\tilde H_r = -\half \alpha (E_i^{(1)}) ^2 + \half K x_i ^2.
\end{equation}
We define canonical momenta from the  Lagrangian (\ref{rm2}). They are
\begin{equation}
p_i = \frac{\pa \tilde L_r}{\pa \dot x_i}= - \alpha B \eps_{ij} E_j ^{(1)}.
\end{equation}
Once again, the introduction of canonical momenta leads to two primary constraints 
\begin{equation}
\tilde \phi_i ^{(0)} = p_i +  \alpha B \eps_{ij} E_j ^{(1)} \approx 0. \label{pcr2}
\end{equation}
Different from (\ref{pbcon}) and  (\ref{pbcon1}), the Poisson brackets between constraints $\tilde \phi_i ^{(0)} \approx 0$ in the present case are vanishing, i.e., $\{ \tilde \phi_i ^{(0)}, \ \tilde \phi_j ^{(0)} \} =0$. It means that there are secondary constraints. Each of the primary constraints (\ref{pcr2}) will lead to secondary constraints. 

By applying the consistency condition to the primary constraints $\tilde \phi_i ^{(0)} \approx 0$, we get
\begin{equation}
\tilde \phi_i ^{(1)} =  \{\tilde \phi_i ^{(0)}, \ H \} = \frac{\alpha k}{r^2} E_i ^{(1)} +K x_i \approx 0. \label{sc}
\end{equation}
We label the primary constraints (\ref{pcr2}) and the secondary constraints (\ref{sc}) in a unified way as $\Phi_I =(\tilde \phi_i ^{(0)}, \ \tilde \phi_i ^{(1)} ), \ I=1,2,3,4$. It can be verified that ${\rm Det } \{ \Phi_I, \ \Phi_J \} \neq 0$. Thus, there are no further constraints and all the constraints $\Phi_I$  are second class. 

It means that when we turn off the electric field $\mathbf E^{(2)}$, the reduced model of (\ref{la20}) does not have  dynamical degrees of freedom. Thus, the electric field $\mathbf E^{(2)}$ plays important roles in producing the FAM: although it does not contribute to the fractional part of the angular momentum directly, the FAM will not appear in the absence of it. 


\section{Conclusions and Remarks}

In this paper, we propose  to simulate anyons by using a trapped cold atom which possesses an induced electric dipole  moment interacting with electromagnetic fields. 
Electromagnetic fields we applied contain a uniform magnetic field and two electric fields.  

We prove that the canonical angular momentum of the model (\ref{la1}) can  only take integer values. 
However,  its  reduced model which is obtained by cooling down the atom to the limit of the  negligibly small kinetic energy  that produces the FAM. The magnitude of the FAM can be modulated by two parameters, i.e.,  the intensity of the applied magnetic field and the  electric field $\mathbf E^{(1)}$. Apart from the fractional part, it is also interesting to observe that the differences between eigenvalues of canonical angular momentum are half integers. It is one of the characteristics of Chern-Simons quantum mechanics. In Ref. \cite{baxter}  the author proposed to realize the Chern-Simons quantum mechanics model by a cold Rydberg atom.  

All the electromagnetic fields play important roles in the simulation of FAM.  
The effect of the electric field $\mathbf E^{(1)}$ is evident since 
the magnitude of the FAM is proportional to the parameter $k$, which is the strength of electric field $\mathbf E^{(1)}$. Roles the electric field $\mathbf E^{(2)}$ played are subtle. At the first glance, the electric field $\mathbf E^{(2)}$ does not contribute to the fractional part of the angular momentum. However, it does influence the results since the FAM will not appear  in the absence of it. 

Besides the contribution to effective mass, roles the magnetic field played will be more transparent if we introduce the effective gauge potentials 
$
A_i ^{\rm eff} = \alpha \eps_{ij} E_j \label{egp}
$
and rewrite the Lagrangian (\ref{la3}) as
\begin{equation}
L = \half M \dot x_i ^2 -  B A_i \dot x_i + \half \alpha E_i ^2 - \half K x_i ^2.  \no
\end{equation}
The interaction term is similar with a charged particle  minimally coupling a gauge field. The magnetic field (\ref{mf}) acts as the coupling strength.  Therefore, the magnetic field not only contributes to the mass of the atom, but also is the coupling strength of the interaction between the atom and the electric fields which is of fundamental importance in  producing the FAM. 

\section*{Acknowledgements}

This work is supported by NSFC with Grant No. 11465006 and partially supported by 20200981-SIP-IPN and the CONACyT under grant No. 288856-CB-2016.


\begin{thebibliography}{99}
\bibitem{note}
This is due to dynamical properties of the Chern-Simons gauge field and in the absence of Maxwell terms.
\end{thebibliography}
\end{document}